\documentclass[12pt,a4paper,aps]{article}
\usepackage{a4}
\usepackage{epsfig}
\usepackage[hmargin=2cm,vmargin=2cm,nohead]{geometry}
\begin{document}

\title{The End of Cheap Uranium} 

\author{
Michael Dittmar\thanks{e-mail:Michael.Dittmar@cern.ch},\\
Institute of Particle Physics,\\ 
ETH, 8093 Zurich, Switzerland\\
\date{June 17, 2011}
}
\maketitle

\begin{abstract}
\noindent
Historic data from many countries demonstrate that on average no more than 50-70\% 
of the uranium in a deposit could be mined. An analysis of more 
recent data from Canada and Australia leads to a mining model with an average deposit extraction
lifetime of $10\pm 2$ years. This simple model provides an accurate description of the extractable 
amount of uranium for the recent mining operations. 

\noindent
Using this model for all larger existing and planned uranium mines up to 2030, a  
global uranium mining peak of at most $58 \pm 4$ ktons around the year 2015 is obtained. 
Thereafter we predict that uranium mine production will decline to at most $54 \pm 5$ ktons by 2025 
and, with the decline steepening, to at most $41 \pm 5$ ktons around 2030.
This amount will not be sufficient to fuel the existing and planned 
nuclear power plants during the next 10-20 years. In fact, we find that it will be difficult 
to avoid supply shortages even under a slow 1\%/year worldwide nuclear energy phase-out scenario 
up to 2025. We thus suggest that a worldwide nuclear energy phase-out is in order. 

\noindent
If such a slow global phase-out is not voluntarily effected, 
the end of the present {\bf cheap uranium} supply situation will be unavoidable. 
The result will be that some countries will simply be unable to afford sufficient uranium fuel at that point, which implies involuntary and perhaps chaotic nuclear phase-outs in those countries involving brownouts, blackouts, and worse.


\end{abstract}


\newpage
\section{Introduction}
Nuclear fission energy in industrial societies is often proposed as a long term 
replacement for the limited fossil fuel resources and as a solution to the environmental 
problems related to their use. 
However, even 50 years after commercial nuclear fission power began, 
nuclear reactors produce less than 14\% of the world's electric energy, 
which itself makes only about 16\% of our final energy demand~\cite{IEA}. 
More than 80\% of the 440 nuclear power plants, with a capacity of 374 GWe~\cite{PRIS},
are operated in the richer OECD countries, where they produce about 21\% of the annual electric 
energy~\cite{IEA}. 
The relatively small nuclear energy contribution today indicates that even a minor transition from fossil to nuclear fuel for generating electric energy over the next 20 to 30 years would require significant increases in the use of nuclear fuel.

Prior to the 2011 Fukushima disaster, mainly China, India and Russia 
had plans for rapid growth of nuclear power during coming decades. 
However, because of the lack of replacement strategies for the aging nuclear reactors in the OECD countries, not even the Word Nuclear Association (WNA) could imagine more than a small worldwide nuclear growth scenario, something on the order of 1-2\%/year\cite{WNAgrowth}. 
Among the many problems related with this small growth scenario is the little discussed 
but fundamental issue of uranium fuel supply~\cite{NEApress}.  

In this paper we present our findings about the future uranium supply.  Our results are obtained 
from a study of deposit depletion profiles from past and present uranium mining. 
Our analysis shows that the existing and planned uranium mines up to 2030 
allow at most an increase of the uranium supply from 54 ktons (54 000 tons) in 2010~\cite{WNAmines} 
to $58 \pm 4$ ktons in 2015. Furthermore, the data indicate that after 2015 production will decline 
by at least 0.5 ktons/year. The annual uranium supply around 2025 and 2030 is thus predicted to 
reach at most $54 \pm 5$ ktons and $41 \pm 5$ ktons respectively. 
These numbers are not even anywhere near the {\it present} global usage, about 68 ktons/year, and imply significant shortages over coming decades.
We thus predict an end of the current situation of cheap uranium and 
a voluntary or forced worldwide nuclear phase-out scenario. 
It is in fact roughly consistent with the new policies,
following the Fukushima accident,
proposed in May 2011 by the governments in Germany and Switzerland.

We start our analysis with countries where uranium mining was stopped or reduced to about 10\% of the past production levels because of depletion (Section 2). The more accurate recent 
mining data from Canada and Australia are used to formulate a simple and accurate mining and 
depletion model (Section 3). In Section 4 this model is applied to the currently operating and planned future 
uranium mines up to 2030.  

\section{Lessons from past uranium mining and depletion}

Significant uranium extraction started after the Second World War~\cite{RB40Y, RB09}.  
Including the year 2010, a total of about 2.5 millions tons of uranium have been mined and about 
2 million tons have been used for electric energy production. 
Most of the remaining 500 ktons are essentially under the control of the military in Russia and the USA. 

Uranium mining between 1945 and 2005 can be divided into three periods. 
The first period (1945-1975) can be associated with the rush to fulfill the military uranium requirements during the nuclear arms race. An extraction peak of almost 50 ktons/year was achieved 
around the year 1959, after which mining declined to about 35 ktons/year between 1965-1975. 
About 750 ktons of uranium were extracted during that period. 

The second period (1975-1990) coincided with the time when many civilian nuclear power plants were planned and constructed. 
This period ended around the year 1990,  
when annual uranium requirements became larger than the annual extraction.  
During this period, uranium mining increased within a few years from 40 ktons to a 
production peak of almost 70 ktons/year around the years 1980/81. A production level 
of more than 60 ktons was maintained between 1978 and 1986
and a total of 1000 ktons were extracted between 1975 and 1990.  

During the third period (1990-2005) the construction of new nuclear power plants 
essentially stopped at a capacity of about 374 GWe, far below the original ambitious 
plans in many countries from the 1970's. During this period and due to depletion and environmental reasons, uranium mining stopped in many productive regions and countries in Europe, 
Africa and North-America. Mining was reduced to an average of 
about 35 ktons/year, well below the uranium demand of 65 ktons/year, and  
a total of 500 ktons were mined.   

During the past five years about 250 ktons of uranium were produced and the
fast rising contribution from Kazakhstan from 4.4 ktons in 2005 to almost 18 ktons in 2010
might be used as an indication that a new production period has started. 
    
\subsection{Uranium depletion I: Europe and Africa}

Uranium mining in Europe ended during the 1990's, and a 
total of about 460 ktons had been extracted when the last mines closed\cite{RB40Y}.
Production reached peak of more than 12 ktons/year in 1976 and maintained a production of 
about 10 ktons/year up to about 1990 from where it steeply declined to   
less than 1000 tons in the year 2000. 

The fact that essentially all of Europe's required 21 ktons/year uranium must now be imported is worth noting since it demonstrates that uranium, like the fossil fuels, is a finite resource that does not somehow magically appear in greater quantities just because demand pushes its price higher. 
As with the fossil fuels, the mining data from Europe show that deposit depletion and production 
declines are unavoidable consequences of finite resources.    


The depletion of mineable uranium in Europe also makes is rather easy to
compare original uranium resource estimates for all major uranium 
producing countries in Europe\cite{IAEAdatabase} with the achieved 
total amount of produced uranium. 
As can be seen from Table 1, mining stopped when 50-70\% of the initial deposit estimates 
had been mined. A similar conclusion can be drawn about the smaller 
but formerly important mines in the D.R. Congo and Gabon\cite{IAEAdatabase}.

The pollution and radioactive remains from these past quick and dirty profitable mining activities leave at least a heavy moral responsibility for those who profited during the past decades from this cheap uranium. 

{\tiny
\begin{table}[htb]
\vspace{0.3cm}
\begin{center}
\begin{tabular}{|c|c|c|c|c|c|}
\hline
country   & demand 2010   & peak production    & initial resource         & extracted              & extracted   \\
               & [ktons]  & [ktons] (year)      & estimate [ktons]       & total [ktons]                          &  fraction    \\
\hline
Germany   &   3.45   &  7.1 (1967)                          & 334.5               &      219.5           & 66\%  \\
Czech Rep. & 0.68    &  3.0 (1960)                         & 233.4               &      109.4                & 47\%  \\
France  &    9.22      &   3.4 (1988)                        & 110.8               &        76.0                        &  69\%  \\
Bulgaria  &   0.28     &   0.65 (1985-1988)                         &    49.1               &       16.4                       & 33\%  \\
Hungary  &  0.30     &   0.6 (1960-1983)                        &    32.8                &       21.1                       & 64\%  \\
Romania  &  0.18     &   2.0 (1956-1958)                         &    37.1               &       18.4                          & 49\%  \\
Spain  &       1.46     &   0.26(1994-2000)                         &    26.4               &         5.0                         & 19\%  \\
\hline
West. Europe  & 21    & 12.3 (1976)  &     $\approx 810$ &    $\approx 460$             & 58\% \\ 
\hline
D.R. Congo                 &  -    & 2.5 (1951)  &   33.5           &    26.5                       & 79\%        \\
Gabon                       &  -    & 1.1 (1979)  &   27.9           &    25.4                      & 91\%        \\

\hline
USA              & 19.4     &    16.8 (1981)  &  1651.8                   &      366.8+472.1$^{*}$   &    51\%      \\
South-Africa &  0.32     &      6.1 (1981) &    444.2                   &      157.4+195.2$^{*}$    &   79\%     \\
\hline
\end{tabular}\vspace{0.1cm}
\caption{Uranium mining data\cite{WNAmines,RB40Y,RB09,IAEAdatabase} from countries 
where mining came to an end many years ago.
The extracted amount of uranium reaches between 50-70\% of the amount 
originally estimated. The data for the USA and South Africa, where the annual mining 
results have been for many years at less than 10\% of their previous peak production years,
are also given. 
As mining has not yet stopped in both countries, the maximal extractable fraction with respect to the 
original amount estimated is obtained  
from the uranium extracted so far and adding their RAR$^{*}$ resources.}
\end{center}
\end{table}
}
 
\subsection{Uranium depletion II: USA and South Africa}

Uranium mining in 2010 in the USA and in South Africa provided 1.7 ktons and 0.6 ktons 
respectively\cite{WNAmines}, namely about 10\% 
compared to their peak production at the 
beginning of the 1980's when 16.8 ktons (USA) 
and more than 6 ktons/year (South Africa) were mined. Including 2010, 
mining operations in the USA have extracted a total of 366.8 ktons and the 
remaining reasonable assured resources, RAR, are given as 472.1 ktons\cite{RB09}.
Both numbers together are much smaller than the original in-place estimate for the known deposits in the USA which was 
1651.8 ktons\cite{IAEAdatabase}. 
For South Africa the total extraction so far is given as 157.4 ktons and the remaining RAR 
resources  are estimated to be 195.2 ktons. 
Again, the sum of both numbers 
is significantly smaller than the original deposit estimate of 444.2 ktons\cite{IAEAdatabase}. 

Moreover, in light of the fact that annual production has declined steeply to roughly 10\% of the peak level that was achieved in the 1980s, the present RAR numbers do not appear to be realistic. Another indication that they are not realistic comes from the very large difference between the present production of the USA and South Africa and the claimed mining capacities in those two countries which are roughly 2.5 and 10 times larger respectively.

It is also important to note that, despite their roughly equal RAR numbers,
uranium production in the USA reaches at most 20\% of that   
from Canada. Similar conclusions about the RAR numbers are obtained when one compares uranium production in South Africa with the currently much larger production of Namibia and Niger.   

\section{A hypothesis about the mining of uranium deposits} 

With a few exceptions, annual mining results from individual mines are publicly available only for the past 
few years. The better documented data from uranium mining centers in 
Saskatchewan (Canada) and Australia are summarized in Table 2\cite{CaAusmines}. 
The data from the long established mining centers show that the mining was actually undertaken 
on several nearby deposits. These deposits were exploited successively 
and in such a way that a relatively stable level of production was sustained over decades. 

{\tiny
\begin{table}[htb]
\vspace{0.3cm}
\begin{center}
\begin{tabular}{|c|c|c|c|c|}
\hline
Name            & initial            & extracted     & plateau   & estimated  \\
(deposit)   & estimate [ktons]      & total [ktons] (years)           & value [ktons]    & total [ktons]   \\
\hline
Rabbit Lake (main)                    & 10-25  & 15.8   (74-84)              & 1.4*              &  $14 \pm 3 $   \\
Collins B                                     & 10-25  & 11.3  (85-91)              & 1.9*              &  $19 \pm  4$  \\
Collins A+D                                & 7.5-15 &   8.6   (94-97)              & 2.7*              &  $26 \pm 6$  \\
Eagle Point                                & 25-50  &  24.9  (92-98 +03-10)  &  2.5-3*          &  $27.5 \pm 6$  \\
\hline 
Rabbit Lake  (all)*                      & 52.5-115 & 60.6 (74-10)      & 4.5*                   &   $86.5 \pm 10 $ \\
\hline
Cluff Lake  (5)                          & 14.5-30        & 10/14.4  (80-92/92-03)   & 1/1.4*  & $ (10+14) \pm 5$   \\
Key Lake   (1)                          & 25-50    & 32 (83-87)   &  6.4* &  $ 64 \pm 13$  \\
Key Lake   (2)                          & 25-50    & 42 (89-01)   &  5.4 &   $ 54\pm 11$ \\
Mc Clean Lake (1)                   & 10-25    & 19.2 (99-10)&  2.35&  $ 23.5 \pm 5$ \\
McArthur (Zone 2)                   & $\approx$ 100 & 73.6 (99-10)& 7.2 & $ 72 \pm 14$   \\
\hline
Ranger (Number 1) & 25-50      & 36.9 (80-95)& 2.8 &   $ 28 \pm 6$ \\
Ranger (Number 3) & $>$ 100  & 55 (97-2010) & 3.8-4.8 & $ 48 \pm 10 $   \\

Olympic Dam          & 300** & 50 (88-98/99-2010)  &1/3 & $ (10+30) \pm 8$   \\
Beverley                   & 10-25 & 6.5 (01-10) & 0.9***  &  $ 9 \pm 2 $  \\
\hline
\end{tabular}\vspace{0.1cm}
\caption{Key data for the main deposits from several well documented recent 
uranium mines and deposits in Saskatchewan/Canada and Australia\cite{CaAusmines}. 
Not all details about the annual individual deposit mining are available, 
and the plateau value is approximated from the average$^{*}$ annual production. 
The amount of the initial estimate$^{**}$ for Olympic Dam takes the entire 
content down to a depth of about 1000 m. So far only the ``surface" layer(s) have been 
touched. The value of 0.9 ktons/year for the plateau of the Beverley$^{***}$ mine 
was obtained only for one year. The 2010 result of less 0.4 ktons indicates that the mine 
will be not achieve anything close to the original plans.  
The last column gives the mining model estimate for the total extraction, obtained from the plateau value 
multiplied 10.   
}
\end{center}
\end{table}
}

Taking the essentially depleted Rabbit Lake (1975 - 2010) mining center 
as a typical example one finds that the uranium mined came from 5 medium to 
high grade deposits ($>$ 1\%), discovered between 1968-1980. 
The larger deposits Rabbit Lake (1975-84) and Collins Bay B zone (1985-91) were totally exploited after 9 and 6 years and the smaller deposits Collins Bay A (1995-97) and D (1995-96) were depleted during 3 and 2 years respectively. Mining at Eagle Point started in 1994, was stopped between 1998-2002 and 
started again afterwards. The remaining uranium in place for closing mining activities 
up to 2015 has a much smaller ore grade (below 1\%) and is estimated, with 
suspect precision, as 8.172 ktons. 

The maximum production at the Rabbit Lake milling facility was about 4.5 ktons/year (1997/98), after  
which it decreased to less than 1.5 ktons/year during the years 2008-2010.
The total production between 1975 and 2010 from the five deposits 
amounts to about 61 ktons. This is on the low side of the 
initial resource estimate, given as somewhere between 52.5 - 115 ktons. 

The McArthur deposit, discovered in 1988, is currently the 
only remaining really productive uranium mining center in Canada. 
The deposits are located about 600 m underground and an estimated average ore grade,15.24\%, is about 100 times larger than those of essentially all other deposits.
The uranium is located in six close deposits, called Zone 1-4, Zone A and B.  
Mine construction started in 1997 and production from the largest deposit, Zone 2,     
began in 1999.  The designed production plateau of 7.2 ktons/year was reached 
in November 2000 and a total of 73.6 ktons have been mined between 2000 and 2010.
The estimates from the 2009 Cameco technical report\cite{TPcameco} about Zone 2 indicate 
that the main body from Zone 2, Panel 1-3, will essentially cease production in the near future. 
Between 2011-2017, production will come from Zone 2,  
Panel 5, with a planned production plateau of about 4.5 ktons/year.   
The results from the first three months of 2011 show a 
production drop of about 35\% when compared with the same period from 2010\cite{Cameco2011}.
This decrease might be a natural consequence of the transition to   
a new Panel (deposit).  

The corresponding data for the still active mines in Australia, Ranger, Olympic Dam and Beverley are 
also presented in Table 2. Perhaps with the exception 
of Olympic Dam, the conclusions about production profiles and mine depletion 
are similar to the one in Canada. 

Uranium from Olympic Dam is defined as one huge very low grade deposit of 0.06\%, lying deep underground. The uranium is distributed over a 3.5x5 km$^{2}$ area and lies between 260-1000 m below the surface.
The huge volume is estimated to contain more than 2 million tons of uranium from which up to 300 ktons are counted as proved ore reserves. Uranium extraction is done in combination with copper and the low grade   
requires an open pit operation. Following the overall extraction profile, about 1 kton/year of uranium was
extracted between 1988 and 1998 and more than 3 ktons/year
after the year 2000. The actual mining could thus be considered as 
operating on deeper and deeper levels similar to the mining of several deposits.

\subsection{The Hypothesis: Deposit exploitation and mining lifetimes}

The data in Table 2 indicate that mining operations seem to be planned such that 
a relatively stable production plateau from a given deposit can be maintained for at least five years.
Assuming that the mining companies use more accurate and partially secret geological information 
about the deposits and design the mining infrastructure in such a way that 
costs are minimized, we propose a working model for the mining of uranium deposits:  

\begin{itemize}
\item The plateau production can be sustained for 
$10 \pm 2$ years. The assumed lifetime uncertainty is guessed from the few years of   
lower extraction during mine start-up and termination.
\item The total amount of extractable uranium 
is approximately the achieved (or for future mines,  the planned) annual plateau value multiplied by 10.     
\end{itemize}

Applying this hypothesis and using the different plateau values, the total 
estimated amount of extractable uranium from the different deposits is given in the last column of Table 2. 
Leaving the three special cases, Collins A+D, Key Lake(1) and Olympic Dam aside, 
the total produced uranium from all mines is about 310 ktons. This is in amazing agreement with our hypothesis which predicts $319 \pm 24$ ktons of extractable uranium. 
 
For the two exceptions, Collins A+D and Key Lake(1), our model predicts much larger  
amounts than achieved. These deposits were only minor when compared to the main deposits nearby
and the short mining periods indicate that the mining of these deposits were probably planned 
such that a smooth transition between the larger deposits was possible. Even though 
our model is so far also in good agreement with the mining at the Olympic Dam deposit, 
because of this somewhat special very low grade deposit,  we prefer to exclude this mine until 
more details become public.

\section{Extraction profiles, the future demand/supply situation}

The above model (hypothesis) can now be applied to the larger currently operating 
uranium mines and the ones which are in a serious planning phase. 
Considering that the achieved plateau values from past operations were usually smaller than the 
ones planned, our model probably overestimates the future production. 
Furthermore, as in the past, 
the planned start-up dates for new mines are uncertain and delays of several years are common.  
It follows that the results from our model should be understood as a likely upper limit for the 
achievable uranium production during the next two decades.
The predicted uranium production for different important mining countries for the next 20 years 
are summarized in Table 3 and a few  
interesting details are discussed below. 

Starting with mining in Canada up to 2030, assured production will come essentially only from 
the McArthur deposit, as large uncertainties are associated with the mining of deposits at Cigar Lake and Midwest. 
The detailed plans for the McArthur deposits\cite{TPcameco} indicate that  
the mining of 4.5 ktons/year from the largest Zone 2 can continue up to 2016.
Extraction from Zone 4 (lower) should start in 2011 and about 2-2.5 ktons/year should be added 
such that the current plateau value of about 7 ktons/year can be maintained until 2016. 
The mining of Zone 4 (upper) part should start around 2016 and should reach a 10 year plateau value of more than 2.5 ktons/year after 2020.  Some time after 2015 the extraction from the smaller deposits in Zone 1 and Zone 3 should also start, adding about 1.5 ktons/year for both of them combined 
during the years 2019 - 2029.
Assuming that the mining from the different Zones take place as planned, 
a continuation of the current annual production of about 7.2 ktons/year can be sustained until 2017.
Around 2017 the production is predicted to drop by 40\% leading to a new production plateau 
of 4.5 ktons/year until 2030. Between 2030-2033, production will end with an annual decrease of roughly 1.5 ktons/year. The plans assume optimistically that about 90 ktons, or about 100\% of the remaining proven and probable reserves, can be extracted during the next 20 years. 

Perhaps the most unlikely detail about this two-decade plan is that the uranium grade 
required to achieve the goals has to increase from an average of 15\% exploited during the last ten years 
to 23\% from 2020-2023 and further to more than 29\% for the remaining years. When one 
considers that after 2020 a large fraction of the uranium has to come from the 
less well understood deposits, the required transition to mining of much higher uranium grade looks 
extremely unlikely, and we predict that real production from the McArthur mine in the new Zones will be significantly lower than the 4.5 ktons/year expected after 2016. 

The Cigar Lake deposit looks in many respects similar to the McArthur deposit.  
However, this deep deposit is located in very poor underground conditions and  
severe flooding incidents have delayed the original mining plans so far by almost ten years.
The latest indications from Cameco assume that mining can start at the end of 2013 and reach 
full production of 7 ktons/year a few years later (2017?). 

According to our model we find that the total extractable amount will be about $70 \pm 14$ 
ktons. This number matches well with the one given for the proven and probable reserves 
of about 80 ktons. It follows that the current mining plans do currently not consider the 
additional 52 ktons of inferred resources.  
We thus predict that this mine will stop production about ten years after the production plateau is 
reached.
 
A few other much smaller deposits in Canada are currently under study. But, even the plans for the most 
advanced Midwest project, with measured and indicated resources
of about 16 ktons, are now rather unclear. 

Assuming that Cigar Lake and Midwest can start around 2017 and 2020 respectively, 
uranium mining in Canada will decline from last year's 9-10 ktons/year to about 7.5 ktons/year 
during the next six years. Once Cigar Lake and Midwest start,    
production can increase again to at most 11 ktons/year up to 2027, whereupon it will decline to 
3 ktons/year or less by 2030.  

For the mines in Australia\cite{AustraliaWNA}
we expect that the decline observed at Ranger, Beverley and the "old" Olympic Dam mine will continue 
during the next few years and production will cease around 2015. 
Perhaps the most interesting project for uranium mining is the Jabolanka deposit in or near 
the world heritage Kakadu National Park. However, this project is currently on hold 
and strong opposition might not allow to open it during the next decades.

The largest future project in Australia (and on the planet) is the mega Olympic Dam 
project\cite{JohnBusby}, a giant open-pit mine, up to 1000 m deep, which might eventually produce 
up to 16 ktons/year. The plans are currently rather badly defined, and we     
assume that an increase to more than 3 ktons/year can at best be imagined after 2020.
Another eight smaller deposits, with a combined reserve estimate of about 80 ktons, 
appear to be in a more serious planning phase and some of them 
might start during the next few years. We assume that the 
mining of these deposits will be stretched over the next decades such that 
in combination with Olympic Dam a total production of about 6 ktons/year can be maintained 
beyond 2030. 

Uranium production in Kazakhstan rose in a spectacular way from about 1 kton/year (1997) and 
4.4 ktons (2005) to make the country the worlds largest uranium producer in 2009.  
The almost 18 ktons produced in 2010 came from the mining of more than 20 smaller deposits
and exceeded the production from Canada and Australia combined.

According to the 2009 Red Book\cite{RB09}, the numbers submitted from this country, 
indicate that this spectacular rise can continue only for a few more years and will decline 
after a peak production of about 28 ktons in 2015 to 24 ktons (2020), 14 ktons (2025), 
12 ktons (2030) and 5-6 ktons in 2035. The latest 2011 announcements from the State-owned mining company, Kazatomprom, are already significantly smaller than the ones from the Red Book 
and mining is now expected to yield 19.6 ktons in 2011 and 20 ktons in 2012. After 2012 mining will stay around the plateau value in the {\it neighborhood} 
of 20 ktons (reaching at most 25 ktons)\cite{Kazatomprom2010}.  

In the absence of precise data from individual mines, we presume that 
a plateau value of $22 \pm 2$ ktons will be maintained until 2015. 
Using the start-up and plateau data summarized in reference \cite{KazakhstanWNA} we find that 
a decline of roughly 1 kton/year for every year after 2015 will be unavoidable.
Mining will decrease accordingly to $17 \pm 2$ ktons around 2020, $12 \pm 2$ ktons in 2025 and to 7 ktons or less in 2030. This predicted decline happens about five years earlier than the one predicted by the Red Book, but given the large uncertainties regarding past and current individual mines in Kazakhstan, the two predictions might be seen as in the same ballpark.

Besides the projects in Canada, Australia and Kazakhstan, five mines 
with more than 1 kton/year production are currently in operation: Rossing and Langer Heinrich in Namibia,
Arlit and Akouta in Niger and Kraznokamensk in Russia.

The Rossing mine, which began production in 1976, is often presented as a proof that uranium can 
be mined efficiently down to grades of 0.03\%. Mining operations cover two large (50-100 ktons and 25-50 ktons) and two smaller deposits (5-10 ktons). Current plans for this mine indicate that production can be maintained up to about 2016 and perhaps up to 2021. 
A more careful analysis of the data from this deposit reveal the special character of this mine. 
According to a 2009 report\cite{IAEAdatabaserossing} the large very low grade deposit contains 
many small clusters of uranium with grades of about 1\%.  
Thus it is most likely that it is mainly these higher-grade clusters that are mined while an overall average grade is reported.
In the absence of better data we assume that the 2010 production value can be sustained until 2020.    

The Langer Heinrich mine started production in 2007, and 1.4 ktons were produced in 2010.
The plateau target is about 2 ktons/year. Following our model,  this 
mine will cease production around 2020 with a plateau value of $1.7 \pm 0.3$ ktons/year.   
 
Several large low-grade deposits are associated with the two older mining centers Arlit and Akouta in Niger.
Production started around 1978 and was increased after 2003.  Both centers together produced about 4.2 ktons in 2010. Taking a plateau value of about 4 ktons/year 
the currently estimated proven reserves allow that full production can be maintained until 
about 2016 and decrease afterwards. The current plateau production 
would thus roughly match the ten-year lifetime model for both mining centers.     

The last operating uranium center in Russia is Kraznokamensk, Eastern Siberia, where  
several deposits have been exploited since decades and the legacy of the related environmental problems 
are detailed in an article by H. H\"ogelsberger \cite{Russiapollution}. 
The remaining total reserve estimates are given as more than 100 ktons and according to a WNA document\cite{Russiamine},  the plateau production of around 3.5 ktons/year coming from several deposits in this area can be maintained or perhaps even slightly increased up to at least 2025.

Concerning other future mines around the planet one finds that seven larger mines, and up to 20 smaller facilities with capacities ranging from a few hundred tons to slightly more than 1 kton/year, are in the planning phase \cite{WNAnewmines}.
Of the seven larger mines, three, with capacities around 5 ktons/year, are planned for Niger (Imourain 2012), Namibia (Husab 2013) and Russia (Elkon 2015), and four, located in Namibia, Jordan, Russia and the Ukraine, will have a combined capacity of about 7-9 ktons/year.
The original start-up plans for these mines have already been delayed by several years.    

In absence of more detailed data we assume that production in Russia can be increased around 
the year 2015 when the production from the Elkon mine will begin\cite{WNArussia}.

For all other future facilities outside of Kazakhstan, Canada, Australia and Russia
we assume that their future production will just allow them to compensate for the 
depletion of the smaller mines operating today and that the combined production in all other countries   
can be kept at today's level of about 17 ktons/year up to 2025-2030.
  
Following our model and combining all countries, we predict a peak uranium production of 
$ 58 \pm 4$ ktons/year (or with 95\% probability less than 66 ktons/year) around the year 2015.  

{\tiny
\begin{table}[htb]
\vspace{0.3cm}
\begin{center}
\begin{tabular}{|c|c|c|c|c|c|}
\hline
Country         & production     &  forecast          & forecast            & forecast       & capacity 
\\
                     & 2010 [ktons]  & 2015 [ktons]    &  2020  [ktons]   &  2025  [ktons]  &  2030  [ktons]   \\
\hline
Kazakhstan   &  17.8       & $22 \pm 2$             &    $17 \pm 2$   &   $12 \pm 2 $       &   $ 7 \pm 2$  \\    
Canada         &    9.8       & $9 \pm 1$             &    $10 \pm 2$   &   $10 \pm 2 $         &   $ 3 \pm 2$  \\    
Australia        &    5.9       & $4 \pm 1$             &    $ 6 \pm 3$   &   $ 6 \pm 3 $         &   $ 6 \pm 3$  \\    
Russia        &    3.6          & $6 \pm 2$             &    $ 6 \pm 2$   &   $ 9 \pm 3 $         &   $ 9 \pm 3$  \\    
all others     &    16.6       & $17 \pm 2$             &    $ 17 \pm 2$   &   $ 17 \pm 2 $         &   $ 17 \pm 2$  \\
\hline
World (max)   &    53.7      & $ 58 \pm 4$             &    $ 56 \pm 5$   &   $ 54 \pm 5 $         &   $ 41 \pm 5$  \\
\hline
\end{tabular}\vspace{0.1cm}
\caption{Uranium production forecast following the Ten-year production model
and some guesswork for the start-up and performance of announced future uranium mines. 
The highest prediction for the year 2015 and the decline should be understood as a 
maximum upper annual production limit during the next 20 years.}
\end{center}
\end{table}
}

\subsection{Uranium demand and other supply estimates} 

Table 4 shows the worldwide uranium demand up to 2030 under a slow nuclear growth (+1\%/year)
and under a nuclear phase-out scenario (-1\%/year). The supply forecast from   
our ten-year production model  and the ones from 
other forecasts are also listed.  
The following considerations are important when one compares the different 
forecasts. 

The WNA 09 forecast\cite{WNAmarket} 
differs from our model mainly in the 
assumption that existing and future mines are assumed to have a lifetime of at least twenty years. 
As a result 
a production peak of 85 ktons/year is envisaged about ten years later and around the year 2025, followed 
by a steep decline to about 70 ktons/year in 2030. 
This prediction could be understood as a warning about the limited uranium supply,  which by 2030 can   
only fuel nuclear reactors with a capacity similar to the ones of today. In any case the long deposit 
lifetime in the WNA 09 model is inconsistent with the data presented in Section 3. 

The 2006 estimate from the Energy Watch Group (EWG)\cite{EWG06} was based on  
the Red Book 2005 RAR and IR (inferred resources) numbers. An upper production 
limit was obtained from the assumption that mining can be increased according to the demand until half of the RAR or at most half of the sum of the RAR+IR resources are used. 
Accordingly, a production peak at latest around the year 2025 is predicted. 
As shown in Section 2, a production profile based on claimed RAR numbers 
is inconsistent with the production in the USA and South Africa.  
We presume that the EWG study would be more consistent with our forecast 
if realistic RAR data were available. 

The largest upper production limit with large uncertainties comes from the  
Red Book capacity scenario\cite{RB09}.  It is acknowledged by the authors of the Red Book that the capacity number given by the different countries is unreliable and much larger than real mining results.
For example, the ratio between the 54 ktons mined in 2010 and the Red Book 09 capacity for 2010 varies between 0.71-0.77. In order to use these numbers nevertheless we presented two 
methods to use the mining capacity numbers for a forecasts\cite{Dittmar09}. 
The RB 09/75 scenario scales all future capacity estimates according to the 2010 ratio 
and with a factor of 0.75. Alternatively the RB 09/N50 scenario assumes that the  
2009 mining result of 50 ktons can be sustained for many years and  
the new mining capacity will be realized with a factor of 0.5. 
Both, the RB 09/75 and the RB 09/N50 scenarios should only be considered as rough 
guesses on how the totally wrong capacity numbers might be used to obtain a approximate 
forecast for upper production limits.
 
When comparing the different forecast, we observe that only our 
simple ten-year lifetime model, by construction, fulfills the condition to be 
consistent with the historic uranium mining data.  
Consequently we conclude that our approach     
provides currently the most realistic upper uranium production limit.  
 
{\tiny
\begin{table}[htb]
\vspace{0.3cm}
\begin{center}
\begin{tabular}{|c|c|c|c|c|c|}
\hline
Scenario         & production   &  forecast          & forecast            & forecast       & capacity \\
                      & 2010 [ktons]     & 2015 [ktons]    &  2020  [ktons]   &  2025  [ktons]  &  2030  [ktons]   \\
\hline
Demand +1\%/year & 68   & 71.5                        & 75                     &   79    & 83    \\
Demand -1\%/year  & 68   & 65                           & 61                     &   58    & 55    \\ 
\hline
this model (max)   &    53.7      & $ 58 \pm 4$             &    $ 56 \pm 5$   &   $ 54 \pm 5 $ &   $ 41 \pm 5$  \\
\hline
WNA 09           &       53.7              & 70                     &   80              &   85               &   70  \\
EWG 06 (RAR-IR)   &    53               & 63-65           &  68-72                  &   70-88          &   65-84          \\
RB 09 (capacity)             &    70-75          & 96-122               & 98-141          &   80 -129   &  75-119  \\
RB 09/75         &    53-57          & 72-91           & 74-106          &   60-97        &  57- 89    \\
RB 09/N50      &     52               & 63-76           & 64-85                  &   55-77           &  53-72           \\
\hline
\end{tabular}\vspace{0.1cm}
\caption{The uranium demand stands either for the demand or the production up to 2030 and 
for a slow $\pm 1\%$/year nuclear growth/phase-out scenario. Our forecast model is the only  
one based on the data from past mining experience and indicates that even with an unchanged 
nuclear capacity a demand of 68 ktons/year uranium shortages can essentially only be 
avoided if the supply from Russian and USA military reserves will continue after 2013.
The other supply scenarios are discussed in the text.
}
\end{center}
\end{table}
}

\section{Summary}

The data about terminated uranium mining 
in different countries and regions demonstrate that on average only 50-70\% of 
initial uranium resource estimates can be extracted.

Using the more precise data about the uranium extraction from recent individual mines 
and deposits in Canada and Australia a depletion model for 
modern uranium mines can be derived. This model states that modern mines minimize the 
extractions costs such that the mining of a given deposit result in 
(1) an effective lifetime of $10 \pm 2$ years and 
(2) the total amount of extractable uranium from a given deposit  
can be approximated by the achieved (or planned) annual plateau value multiplied by 10.     

This model is applied to existing and planned uranium mines
and an upper production limit for 
uranium extraction in different countries and for the entire planet is obtained up to 2030. 
In detail we find that:  
\begin{itemize}
\item A production decline from essentially all mines operating on particular deposits is unavoidable   
during the present decade.  
\item This decline can only be partially compensated by the planned new mines.       
\item Assuming that all new uranium mines can be opened as planned,  
annual mining will be increased from the 2010 level of 54 ktons to about $58 \pm 4$ ktons in 2015.
\item After 2015 uranium mining will decline by about 0.5 ktons/year up to 2025 
and much faster thereafter. The resulting maximal annual production is predicted as 
$56 \pm 5$ ktons (2020), $54 \pm 5$ ktons (2025) 
and $41 \pm 5$ ktons (2030).
\end{itemize}

Assuming that the demand side will be increased by 1\% annually, we 
predict both shortages of uranium and (inflation-adjusted) price hikes within the next five years.

A way to delay a supply crunch until 2025 could be a voluntary nuclear energy phase-out in many countries. Such a phase-out appeared to be very unlikely at the beginning of 2011, 
but the recent accident in the Japanese Fukushima nuclear power complex could lead to totally 
different prospects.  

Another alternative to avoid shortfalls during this decade  
would be a ``wider" opening of the still sizable quantities of the military uranium reserves 
from the USA and Russia especially after 2013.  
Although any such increases involve political issues that clearly go beyond the scope of our analysis, 
this source too is obviously subject to depletion over time.

Therefore, assuming that a global slow phase-out scenario will not be chosen on a voluntary basis, 
we predict that the end of the cheap uranium supply will result in a chaotic 
phase-out scenario with price explosions, supply shortages and blackouts in many countries. \\

\noindent
{\bf \large Acknowledgments\\}

\normalsize
This analysis is a result of many, mostly unfruitful, discussions with pro nuclear enthusiasts  
about uranium being a finite resource like fossil fuels and about 
the difference between actual and future uranium mining and the exploitable amounts 
of uranium in the earth crust. 

Even though the results and views expressed in this paper are from the author alone, 
I would like to thank several friends, colleagues and students who took the trouble 
to discuss the question of uranium resources with me. They all helped me to develop this simple model
and complete this analysis into its present form.   
I would like to thank especially W. Tamblyn, F. Span\`o and C. Campbell 
for many valuable suggestions and for the careful reading of the paper draft. 


\end{document}